\documentclass[reprint,amssymb,showpacs,preprintnumbers,aps,pre,sort&compress,numbers
]{revtex4-2}
\usepackage{amsmath} 
\newcommand{\vect}[1]{\boldsymbol{#1}} 

\usepackage{graphicx}
\usepackage{bm}
\usepackage{etoolbox} 
\usepackage{lipsum} 
\usepackage{hyperref}
\usepackage[capitalize]{cleveref}
\usepackage{xcolor}
\usepackage{float}

\makeatletter
\newcommand*{\rom}[1]{\expandafter\@slowromancap\romannumeral #1@}
\makeatother

\begin{document}


\title{A simple model for self-propulsion of microdroplets in 
surfactant solution}

\author{Swarnak Ray}
\author{Arun Roy}
\email{aroy@rri.res.in}
\affiliation{%
 Soft Condensed Matter Group, Raman Research Institute, Bangalore 560080, India
}%


\date{\today}

\begin{abstract}
We propose a simple active hydrodynamic model for the self-propulsion of a
liquid droplet suspended in micellar solutions. The self-propulsion of the droplet occurs by spontaneous breaking of isotropic symmetry and is studied using both analytical and numerical methods. The emergence of self-propulsion arises from the slow dissolution of the inner fluid into the outer micellar solution as filled micelles. We propose that the surface generation of filled micelles is the dominant reason for the self-propulsion of the droplet. The flow instability is due to the Marangoni stress generated by the non-uniform distribution of the surfactant molecules on the droplet interface. In our model, the driving parameter of the instability is the excess surfactant concentration above the critical micellar concentration which directly correlates with the experimental observations.  We consider various low-order modes of flow instability and show that the first mode becomes unstable through a supercritical bifurcation and is the only mode contributing to the swimming of the droplet. The flow fields around the droplet for these modes and their combined effects are also discussed.  
\end{abstract}
\maketitle

\section{Introduction}\label{introduction}
A common example of an artificial micro-swimmer is a liquid droplet dispersed in another immiscible fluid and propelled by self-generated Marangoni flow. This requires the creation of non-uniformity of surface tension which is maintained by a non-uniform distribution of surface active species on the interface. One way of creating such surface tension gradients is chemical reactions \cite{banno2012ph, ban2013ph, kitahata2011sp, Kasuo2019, Suematsu2021,
Suematsu2016, Suematsu2019, Thutupalli2011, thutupalli2013tuning, Schmitt2013, Yoshinaga2012}. Another way of generating surface tension gradients is micellar solubilization in which a drop of one fluid slowly dissolves in a micellar solution of another fluid by forming filled micelles. Earliest reports of spontaneously generated convective flow fields by such solubilizing droplets can be found in \cite{chen1997rates, chen1998rates, pena2006, Peddireddy2012}. But the authors did not perform a detailed study of self-propulsion. In recent years, several studies have focused on droplet motion due to micellar solubilization \cite{Izri2014, Herminghaus2014, Jin2017, Kruger2016, Suga2018, Moerman2017, izzet2020, Suda2021, Dwivedi2021, Hokmabad2021, Yamamoto2017, Castonguay2023}. Some of these systems involve water droplets in solutions of a non-ionic surfactant in an organic oil \cite{Izri2014, Suda2021}, while some others involve oil droplets suspended in aqueous solutions of ionic surfactants \cite{Moerman2017, izzet2020, Herminghaus2014, Jin2017, Kruger2016, Suga2018, Dwivedi2021, Hokmabad2021, Yamamoto2017}. In all of these systems, it has been found that the drops simultaneously exhibit self-propulsion and dissolution above a sharp threshold total concentration of the surfactant in the outer fluid which is much greater than the critical micellar concentration (CMC). It has also been found that  
the inner fluid can be in the isotropic \cite{Moerman2017, izzet2020, Jin2021}, nematic \cite{Herminghaus2014, Jin2017, Kruger2016, Suga2018, Dwivedi2021} or cholesteric phases \cite{Yamamoto2017} while self-propulsion of smectic droplets has not been reported. It was also shown using fluorescence microscopy that the droplets leave behind a trail of filled micelles \cite{Jin2017, Hokmabad2021}.  

There are several models aiming to explain the self-propulsion of such droplets by proposing mechanisms for sustaining the non-uniform distribution of surfactants. Since there is no inherent asymmetry in such systems, one relies on the spontaneous symmetry breaking of isotropic surfactant distribution.  Herminghaus et. al \cite{Herminghaus2014} proposed that the interface region of the droplet acts as a sink (source) for empty (filled) micelles. This leads to a radial gradient of the density of empty micelles in the steady state. This gradient plays the role of a driving parameter in their model for self-propulsion. 
The authors showed that in systems where the empty micelles collect solute molecules from the bulk, increasing the empty micelle gradient can
result in increased surfactant concentration at the interface. This effect combined with small perturbations of droplet flow fields can lead to the desired self-propulsion of the droplet. However, the assumed far field gradient of the empty micelles in the outer fluids is not expected for such micrometer-sized droplets. In an attempt to formulate a generalized model, Morozov et al. \cite{Morozov2019nonlinear, Morozov2019self} treated the droplet interface as a sink for monomers and assumed a fixed flux condition at the interface. Assuming a  characteristic velocity scale associated with such a fixed flux, they formulated an intrinsic Peclet number (Pe) defined as the ratio of characteristic strength of advection and diffusion. They showed that beyond a critical value of this Peclet number, the nonlinear coupling between concentration and velocity field leads to self-propulsion. However, the mechanism of this characteristic fluid flow proportional to the influx of monomers is not clear. More recently Morozov et al. \cite{Morozov2020} attempted to treat the droplet interface as a source for swollen micelles and developed a similar model. However, the velocity scale used in defining Pe was introduced somewhat artificially in the model. Izzet et al. \cite{izzet2020} showed that a radial gradient of surfactant monomers in the vicinity of the droplets can exist because of the possible lowering of CMC due to the presence of the small number of oil molecules dissolved near the interface. An infinitesimal perturbation in the droplet velocity can lead to anisotropy in surfactant concentration on the droplet interface, inducing a Marangoni flow, and at high enough dissolution rates, the droplet can propel itself.

Most of the existing models take into account only the variation of surfactant concentration outside the droplet \cite{Izri2014, Herminghaus2014, Morozov2019self, Morozov2019nonlinear, Morozov2020} as the mechanism of self-propulsion. One key aspect of such systems found experimentally is that droplet motion can only occur above a certain critical value of surfactant concentration in the bulk far exceeding the CMC. The excess concentration above the CMC produces micelles that are in a dynamic equilibrium with the monomers in the fluid. Therefore the monomer concentration in the bulk is expected to remain close to the CMC value for any deficiencies in monomer concentration should be quickly replenished by the dissociation of empty micelles. Only one of the available models takes this factor into account \cite{Morozov2020}. This model considers the transport of adsorbed monomers at the interface and assumes an explicit form of the filled micelle production rate from the interface. However, it still relies on symmetry breaking of species concentration in the bulk and does not provide a clear relation between activity and total surfactant concentration.
The present study aims to provide a minimal model that directly correlates the onset of self-propulsion beyond a total bulk concentration of the surfactant and to show that self-propulsion could be achieved through interfacial processes alone.

In section \ref{Model}, we describe the geometry of the model system used in our calculations and the mathematical model developed to account for the self-propulsion of a droplet. The linear stability analysis of the model equations is discussed in section \ref{Linear Stability Analysis}. In section \ref{Nonlinear numerical analysis}, we describe the numerical methods used to solve the nonlinear transport equations. The results and conclusions are given in sections \ref{Results} and \ref{Conclusion}, respectively.

\section{Model}\label{Model}
The physical system consists of a swimming droplet of radius $a$ slowly
dissolving into a surfactant solution. The droplet interface is covered 
with surfactant molecules that can be transported along the interface 
due to diffusion and advection. We assume that the solubility of the
surfactant molecules in the droplet's inner fluid is negligible and they 
mostly exist at the interface and the outer fluid. Surfactant molecules 
can exist in three forms in the outer fluid viz. as monomers with
concentration denoted by $C_1$, as empty micelles, and as filled micelles 
that have acquired some molecules of the dissolving inner fluid.  
Both the inner and outer fluids are assumed to be Newtonian and
incompressible in our model. The droplet is assumed to be moving in 
the outer fluid of an infinite extent with no externally imposed flow. 
The density and viscosity of the inner(outer) fluids are homogeneous 
and denoted as $\tilde{\rho}(\rho$) and $\tilde{\mu}(\mu)$, respectively.
Since the Reynolds number ($Re$) associated with these systems is often much less than unity, the inertia of the fluids and the drop is negligible. Hence the flow fields satisfy the Stokes equations. The flow fields are subjected to kinematic, dynamic, and stress balance conditions at the droplet interface. These boundary conditions along with the surfactant transport equation are used to solve for the velocity fields and the surfactant distribution on the droplet interface.

There can be several different mechanisms for the solubilization of an 
oil/water droplet into a micellar solution. (a) It is possible that empty 
micelles directly collide with the droplet interface and collect solute 
molecules and diffuse into the bulk. (b) The empty micelles may acquire 
individual solute molecules from a diffused layer around the droplet 
near its interface. (c) There may be direct emission of solute-filled 
micelles from the droplet interface. The first possibility is not 
expected to be operative for systems with ionic surfactant molecules 
due to the electrostatic repulsion between the micelles and the droplet 
interface. The diffused layer thickness in the second process is also 
expected to be small for low solubility of the inner solute molecules in the outer liquid and surfactant concentrations in the outer liquid well above the CMC. For surfactant concentrations well above the CMC, the large number of empty micelles present in the outer liquid take away the solute molecules from the diffused layer and reduce it to a negligible thickness at a steady state. Hence the self-propulsion due to this process is unlikely. We assume that the spontaneous emission of filled micelles, from the interfacial monolayer of adsorbed surfactant molecules, is the dominant process contributing to the solubilization in our model. The rate of emission of filled micelles is postulated to be proportional to excess surfactant concentration $C_e = (C_{tot} - C_{m})$, where $C_{tot}$ is the total surfactant concentration and $C_m$ is the critical micellar concentration, respectively. The excess surfactant concentration $C_e$ increases the propensity of holding the filled micelles in the outer fluid. This emission is expected to decrease the average interfacial surfactant concentration compared to that of a non-solubilizing drop in a micellar solution. We note that minute perturbation of surfactant distribution leads to small amplitude flow fields near the droplet interface. We propose that in regions of negative surface divergence of these flow fields, there is compression of the monolayer which in turn facilitates the formation of filled micelles from those regions. On the other hand, in regions of positive surface divergence, the stretching of the monolayer hinders the emission of swollen micelles. It is found that the droplet dissolution rates are usually linear which hints that it could be a surface-dominated process \cite{Izri2014, Suga2018, Peddireddy2012}.
\begin{figure}[!htb]
\includegraphics[width=1.0\linewidth]{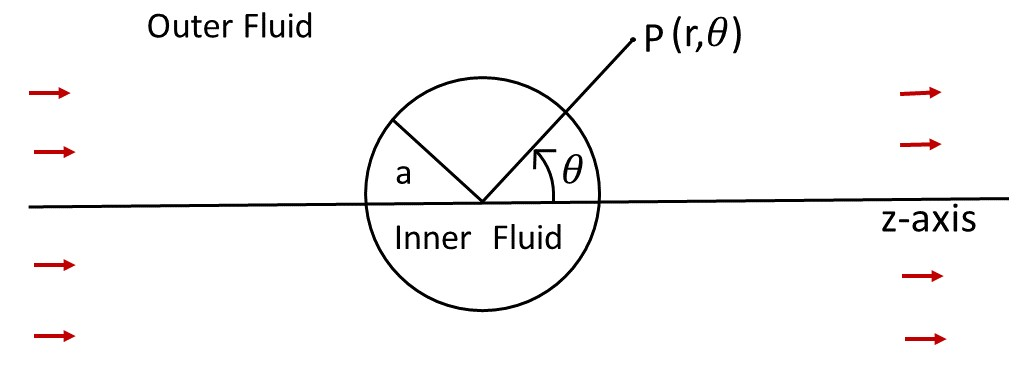}
\caption{Schematic diagram of a swimming droplet of radius $a$ in an outer fluid with coordinates system used to denote a general point P. The red arrows depict the far field velocity of the outer fluid in the droplet rest frame.}
\label{spframe}
\end{figure}
   
   The model equations are made dimensionless by performing the following transformations of the relevant variables.
The radial distance is measured in units of the droplet radius $a$ giving the dimensionless form $r^{*} = r/a$. The dimensionless time $t^{*}=t/( \frac{a^2}{D_s})$, where $D_s$ is the molecular diffusivity of surfactant molecules at the interface.
The dimensionless interfacial surfactant concentration, bulk monomer concentration, and bulk excess surfactant concentration are defined as 
$\Gamma^{*}=\Gamma/ \Gamma_m$, $C^{*}_1= C_1/C_m$, $C^{*}_e = C_e/C_m$ respectively, where $\Gamma_m$ is the maximum possible interface concentration.
The dimensionless surface tension $\sigma^{*}=\sigma/ \sigma_{0}$, 
where $\sigma_{0} $ is the surface tension of a clean interface.
The fluid velocity and pressure are made dimensionless as
$\vect u^{*}= \vect u/(\frac{D_s}{a})$ and $p^{*}= p/(\frac{\mu D_s}{a^2})$. 
The other parameters of the model appearing in eqn.~(\ref{eq:interfacial}) below are made dimensionless using 
$k_a^* = k_a a^2 C_m/\Gamma_m$, $k_d^* = k_d a^2/D_s$, $ e_1^* = e_1 C_m a^2/(D_s\Gamma_m)$ , $e_2^* = e_2 C_m/\Gamma_m$.
For convenience, we henceforth denote the dimensionless parameters and variables without the superscript star.

In the dimensionless form the momentum transport and continuity equations 
for the incompressible outer/inner fluids for low Reynolds number 
can be written as,
\begin{align}
\vect\nabla^2\vect u = \vect\nabla p \: ;\: \vect\nabla\cdot\vect u = 0  \label{eq:Stokes_out}\\
\nu\vect\nabla^2\tilde{\vect u} = \vect\nabla \tilde{p} \:;\: \vect\nabla\cdot  \tilde{\vect u} = 0 
\label{eq:Stokes_in}
\end{align}
where, \{$\vect u,p$\}(\{$\tilde{\vect u},\tilde p$\}) represent the
dimensionless velocity and pressure fields of outer(inner) fluids respectively. All the bulk material properties of the fluids are taken to be constant and the gravitational effects are negligible. The self-propulsion of the droplet along a certain direction occurs with the spontaneous breaking of the isotropic symmetry.  We assume the flow field around the droplet is axisymmetric and solve the equations in the droplet rest frame using a spherical polar coordinate system as shown in fig.~\ref{spframe}.
Without loss of generality, the droplet is assumed to be moving along the negative z-axis in the lab frame, so that in the droplet rest frame the far-field velocity takes the form,
\begin{equation}
\vect u\to U \hat{z}
\label{eq:far_field_boundary_condition}
\end{equation}
where $U$ is the magnitude of droplet velocity in the lab frame and $\hat z$ is
the unit vector along the polar axis.  The dimensionless boundary conditions at the droplet surface, $r=1$ can be written as,
\begin{equation}
 u_r = \tilde u_r =0 \: ,u_{\theta} = \tilde u_{\theta}
 \label{eq:interface_boundary_condition}   
\end{equation}
which represent the vanishing of the normal component of velocity due to the impenetrability of the interface and the continuity of the tangential component of velocity, respectively. We also neglect any small shape changes in the droplet.

The dimensionless form of the advection-diffusion equation for the interfacial surfactant concentration $\Gamma$ can be written as,
\begin{equation}
\begin{split}
 \frac{\partial \Gamma}{\partial t}+\vect{\nabla_s}\cdot(\vect{u_s}\Gamma)=\vect{\nabla}_s^2\Gamma +k_a C_1(1-\Gamma)-k_d\Gamma\\
 -(e_1-e_2 \vect{\nabla_s}\cdot\vect{u_s})  C_e 
\end{split}
\label{eq:interfacial}
\end{equation}
where the terms on the left-hand side of eqn.~(\ref{eq:interfacial}) represent the explicit time derivative and advection of surfactants, respectively. The operator $\vect\nabla_s$ represents the surface gradient on the droplet interface. The first term on the right-hand side represents the molecular diffusion of adsorbed surfactants on the interface. The second (third) 
term on the right-hand side represents the adsorption (desorption) of 
monomers at the interface from(to) the outer fluid with the dimensionless rate coefficient $k_a$($k_d$). The last term in eqn.~(\ref{eq:interfacial}) takes into account the spontaneous emission of filled micelles from the interface. The first part of this term with
coefficient $e_1$ represents an isotropic emission independent of flow and the other part with coefficient $e_2$ represents the emission contribution depending on the flow as discussed earlier. We assume that the bulk monomer concentration $C_1$ remains homogeneous and constant at $C_m$.

For simplicity, we consider a linear relationship between surface 
tension and interfacial surfactant concentration as
$\sigma(\Gamma)= 1 -\frac{ R T \Gamma_{m}}{\sigma_0} \Gamma$, where $R$ is the ideal gas constant and $T$ is the absolute temperature. 
The tangential stress component due to the Marangoni effect is 
discontinuous across the interface whenever $\vect\nabla_s\sigma$ is non-zero and this boundary condition
in the spherical polar coordinate frame can be written as,
\begin{equation}
 \nu \big(-\frac{\tilde u_{\theta}}{r}+\frac{\partial \tilde u_{\theta} }{\partial r}\big) -\big(-\frac{u_{\theta}}{r}+\frac{\partial u_{\theta} }{\partial r}\big)
 = -M \frac{\partial \Gamma}{\partial \theta}    
\label{eq:stress_balance}
\end{equation}
where  $M=\frac{RT \Gamma_m a}{D_s \mu}$ is the Marangoni number and 
$\nu = \frac{\tilde{\mu}}{\mu}$ is the ratio of the dynamic viscosities
of the inner and outer fluids, respectively.

The hydrodynamic equations given by eqn.~(\ref{eq:Stokes_out}) and eqn.~(\ref{eq:Stokes_in}) under axisymmetric conditions can be solved using a stream function formulation with the superposition of different orthogonal modes \cite{Leven1976, Leal2007}. Noting that there is no external body force acting on the droplet, the solutions for the radial and tangential components of the fluid velocity field in the droplet rest frame can be written as \cite{Morozov2019self, Morozov2019nonlinear, Morozov2020},
\begin{widetext}
\begin{align}
u_r &= U\bigg(1-\frac{1}{r^3}\bigg)\eta + \sum_{n=2}^{\infty} \alpha_n n(n+1)\bigg(r^{-n-2}-r^{-n}\bigg)P_{n}(\eta)\label{eq:u_r_out}\\
u_{\theta} &= -U\bigg(1+\frac{1}{2r^3}\bigg)\sqrt{1-\eta^2}  +
       \sum_{n=2}^{\infty} \alpha_n n(n+1)\bigg((2-n)r^{-n}+nr^{-n-2} \bigg)\frac{G_{n+1}(\eta)}{\sqrt{1-\eta^2} }\label{eq:u_theta_out}\\
\tilde u_r &= -\frac{3U}{2}\bigg(1-r^2\bigg) \eta - \sum_{n=2}^{\infty} \alpha_n n(n+1)\bigg(r^{n+1}-r^{n-1}\bigg)P_n(\eta)\label{eq:u_r_in}\\
\tilde u_{\theta} &= \frac{3U}{2}\big(1-2r^2\big) \sqrt{1-\eta^2}+ 
       \sum_{n=2}^{\infty} \alpha_n n(n+1)\bigg((n+3)r^{n+1}-(n+1)r^{n-1} \bigg) \frac{G_{n+1}(\eta)}{\sqrt{1-\eta^2}}\label{eq:u_theta_in}
\end{align}
\end{widetext}

where $\eta = \cos\theta$ and the functions $P_n(\eta)$, $G_n(\eta)$ are
the Legendre polynomial 
of degree n and the Gegenbauer polynomial of order n and 
degree $-\frac{1}{2}$, respectively.
Substituting these expressions for the velocity fields in the stress balance 
condition Eq.~(\ref{eq:stress_balance}) and using the orthogonality 
properties of the Legendre and the Gegenbauer polynomials,
the far-field flow speed can be written as,
\begin{equation}
     U = \frac{M}{3\nu+2} \int_{0}^{\pi} \frac{\partial \Gamma}{\partial \theta}\; G_2\,d\theta
\label{eq:velocity_expression}
\end{equation}
On the other hand, the flow amplitudes $\alpha_n$ of the higher order modes for $n\geq 2$ can be found as,
\begin{equation}
    \alpha_n = -\frac{M}{4(\nu+1)}\int_{0}^{\pi}\frac{\partial \Gamma}{\partial \theta}\; G_{n+1}\,d\theta
    \label{eq:mode_expression}
\end{equation}

\section{Linear Stability Analysis }\label{Linear Stability Analysis}
The linear stability analysis was performed on the nonlinear equations of the model to determine the threshold value of the control parameter above which the reference state becomes unstable. In the reference state, the droplet has a uniform distribution of surfactants on the interface with zero flow in the inner and outer fluids. The linear stability analysis was carried out by introducing small perturbation of order $\epsilon$ to the dependent variables from their
reference values as,
\begin{equation}
\begin{aligned}
    \vect u &= \vect 0 + \epsilon \vect u \\
    \vect{ \tilde{u}}&=\vect 0 +\epsilon \vect{ \tilde{u}}\\
      \Gamma &= \Gamma_0 + \epsilon \Gamma_1 
    \end{aligned}
    \label{eq:form_of_perturbation}
\end{equation}
The different orders of approximation can be obtained by substituting the variables from eqn.~(\ref{eq:form_of_perturbation}) into the equations~(\ref{eq:Stokes_out}) - (\ref{eq:stress_balance}) and collecting the terms of the same powers of 
$\epsilon$. The zeroth order approximation gives the uniform surfactant concentration on the droplet interface in the reference state as,
\begin{equation}
    \Gamma_0 = \frac{k_a-e_1 C_e}{k}  
    \label{eq:reference_state_concentration}
\end{equation}
where $k = k_a+k_d$. In the first order 
approximation, $\Gamma_1$ satisfies the transport equation,
\begin{equation}
   \frac{\partial \Gamma_1}{\partial t} = \vect \nabla^2 \Gamma_1 - k \Gamma_1 + \vect \nabla_s . \vect u_s(e_2 C_e-\Gamma_0) 
   \label{eq:linear_equation}
\end{equation}
and the flow velocities in the fluids satisfy Stokes equations. Then the solutions to the fluid velocity components can be written in the form of 
eqns.~(\ref{eq:u_r_out}) - (\ref{eq:u_theta_in}). Assuming $\Gamma_1$ can be expanded as 
\begin{equation}
    \Gamma_1 = \sum_{n=1}^{\infty}b_n(t)P_n(\cos\theta),  
    \label{eq:Gamma1_exp}
\end{equation}
we solve the resulting time evolution equations for the mode amplitudes $b_n(t)$. Then using 
eqn.~(\ref{eq:velocity_expression}) and eqn.~(\ref{eq:mode_expression}),
the amplitudes for the flow velocities can be written as,
\begin{equation}
\begin{aligned}
 U &= -\frac{2Mb_1}{3(3\nu+2)} \\
 \alpha_n &= \frac{Mb_n}{2(\nu+1)(2n+1)} \textrm{ for } n\geq 2 ,
\end{aligned}
 \label{eq:relation_between_amplitude_and_concentration}
\end{equation}
For the first mode ($n=1$), the amplitude $b_1(t)$ satisfies,
\begin{equation}
    \frac{d b_1}{d t} = \bigg [\frac{2M}{3\nu + 2}(e_2 C_e - \Gamma_0) - (2+k)\bigg ] b_1 
    \label{eq:evolution_of_first_mode}
\end{equation}
which gives $b_1\propto e^{\lambda_1 t}$ with the growth exponent
\begin{equation}
    \lambda_1 = \frac{2 M}{3 \nu + 2}(e_2 C_e - \Gamma_0) - (2+k)  
    \label{eq:growth_rate_1} 
\end{equation}
For $\lambda_1 > 0$, the reference motionless state becomes unstable to the swimming mode ($n=1$) giving the threshold excess concentration as
\begin{equation}
    C_{e1} = \frac{k(2+k)(3\nu+2)+2Mk_a}{2M(ke_2+e_1)}.
    \label{eq:first_threshold}
\end{equation}
Similarly for higher order modes with $n\geq 2$, 
the time evolution equation of the mode amplitude $b_n(t)$ can be
written as,  
\begin{equation}
\frac{d b_n}{d t} = \bigg [\frac{Mn(n+1)}{(\nu+1)(2n+1)}(e_2C_e-\Gamma_0)-\{n(n+1)+k\}\bigg ] b_n
\label{eq:evolution_of_higher_modes}
\end{equation}
and the threshold excess concentration above which the n-th mode becomes
unstable is given by,
\begin{equation}
C_{en} = \frac{k\{n(n+1)+k\}(\nu+1)(2n+1)+k_aMn(n+1)}{n(n+1)M(ke_2+e_1)}
\label{eq:higher_threshold}
\end{equation}
It should be noted that only mode~1 gives rise to the net propulsion of the droplet. The higher-order modes though produce flow around the droplet, do not give rise to the net propulsion of the droplet as discussed below.

\section{Nonlinear numerical analysis}\label{Nonlinear numerical analysis}
The linear stability analysis gives the threshold values of the excess concentration $C_e$ above which reference motionless state becomes unstable to different instability modes. Above the threshold, these modes initially grow exponentially with time but become saturated at long times due to the non-linear effects. Hence the full non-linear model equations need to be solved to study the long-time behaviour of these modes. The non-linear surfactant transport equation was solved numerically using a forward time central space (FTCS) finite difference scheme to find the saturation values of the mode amplitudes above the threshold.
In this numerical method, the surface velocity field given by eqn.~(\ref{eq:u_theta_out}) for modes under consideration was substituted in 
eqn.~(\ref{eq:interfacial}). The resultant equation was discretized for the spatial derivatives using a second-order central difference scheme and the time integration was performed using the forward Euler method. Because of the assumed axisymmetry of the problem,  eqn.~(\ref{eq:interfacial}) can be solved in the half-space $0\leq\theta\leq\pi$. The condition of axisymmetry also requires that the diffusive flux of the surfactants on the drop interface be zero at the 
poles $\theta = 0$ and $\theta =\pi$. In our numerical scheme, this was accomplished by using ghost points outside the range of $\theta$. 
L'Hospital's rule was used to remove the singularity in the diffusive term ($\vect{\nabla}_s^2\Gamma$) at the poles. For the first mode, the solution was advanced in time assuming an initial form of the interfacial surfactant concentration 
$\Gamma= \Gamma_0 -\epsilon_1 P_1(\cos{\theta})$, where $\epsilon_1$ is a small amplitude perturbation to the uniform concentration. The solution was evolved until the surface concentration reached a steady distribution and the velocity amplitudes reached a saturation value. The same method is used for mode 2 and for
the combined mode with suitable forms for the initial perturbations.  
To test the accuracy
of the above method, we solved eqn. (\ref{eq:interfacial}) using one other scheme:  Forward Euler in time, explicit treatment of advection term, and implicit treatment of diffusion term using Crank-Nicholson method with a second-order central difference for spatial derivatives. Both methods gave very similar results with negligible differences in the solutions. All the computations for different values of the parameters were carried out using the relatively faster FTCS method. 

\section{Results}\label{Results}
The model equations were solved numerically considering the low-order modes of velocity fields which are the dominant modes controlling the hydrodynamic signature of these micro-swimmers. The following values of the 
dimensionless model parameters are used in the calculations: 
$M = 24775$, $\nu = 60.0$, $k_a = 0.4$, $k_d = 0.0$,
$e_1 = 0.0048$, $e_2 = 0.008$. Below we discuss the results obtained for
different instability modes.
\subsection{Mode 1}
For the first mode, the non-uniform surfactant distribution is given by the Legendre polynomial of degree one and has vectorial symmetry. The velocity fields corresponding to this mode are given by the first term in eqns.~(\ref{eq:u_r_out}) - (\ref{eq:u_theta_in}). This mode gives rise to net self-propulsion of the droplet consistent with its vectorial symmetry.  The eqn.~(\ref{eq:first_threshold}) shows the expression for the threshold value 
$C_{e1}$ of the driving parameter from the linear stability analysis which
agrees with the non-linear analysis for the above values of the model parameters.
\begin{figure}[!htb]
  \centering
  \includegraphics[width=1.0\linewidth]{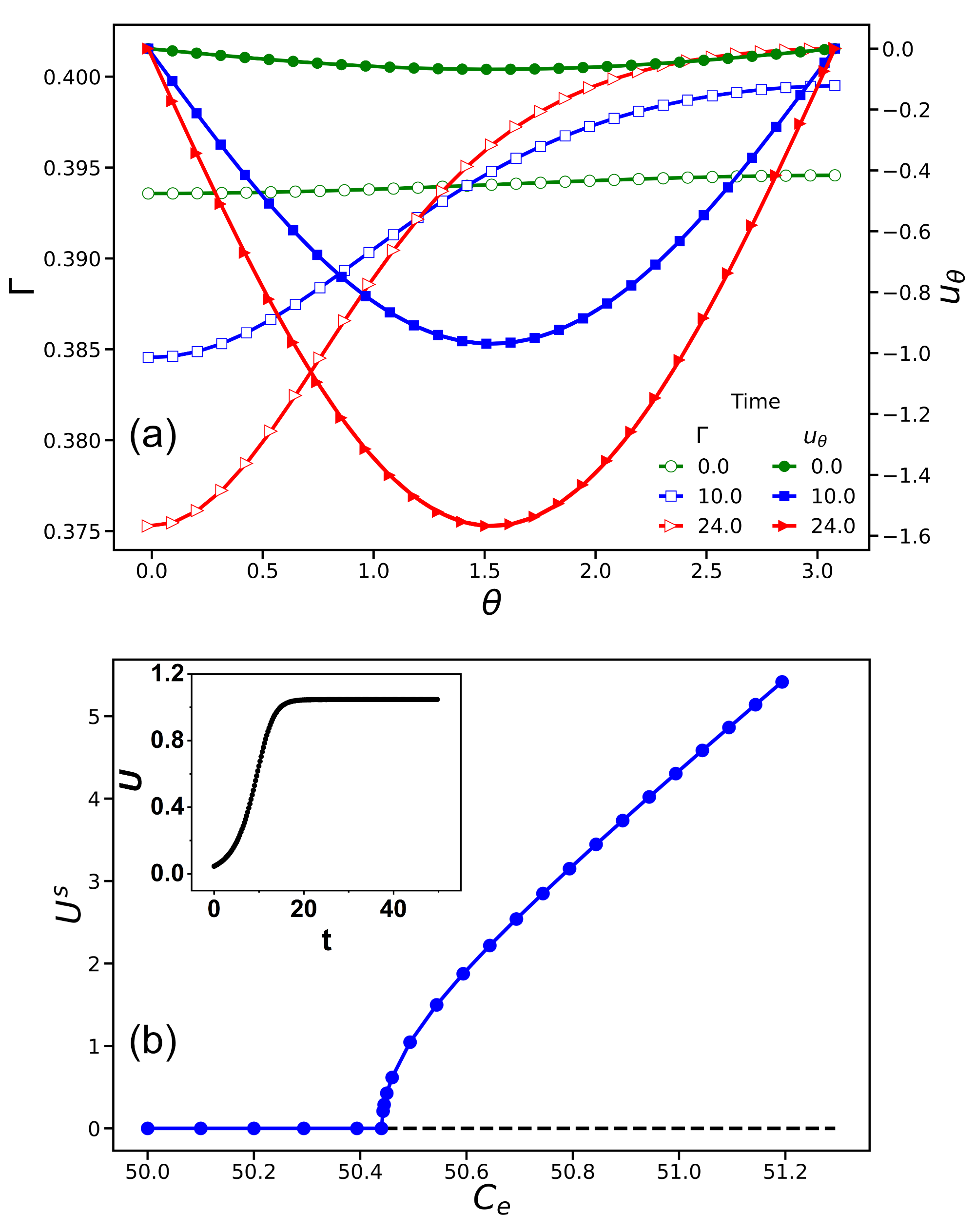}
  \caption{(a) The time evolution of surfactant concentration $\Gamma$ and surface velocity $u_s$ at the droplet interface at three different instants of time for $C_e = 50.493$ corresponding to the first mode of instability. The threshold value of the control parameter for this mode is $C_{e1} = 50.440$. (b) The variation of the steady-state droplet propulsion speed $U^s$ with the control parameter $C_e$. The $U^s$ grows continuously from zero above the threshold indicating a supercritical bifurcation for this mode. The inset depicts the time evolution of $U$ for $C_e = 50.493$ showing the saturation at long times.}
  \label{fig:Mode_1}
\end{figure} 
\begin{figure}[!htb]
  \includegraphics[width = 0.7\linewidth]{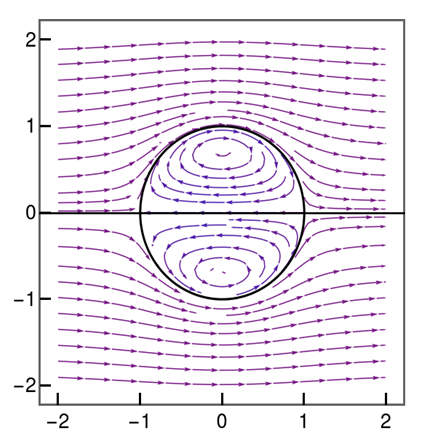}
  \caption{The numerically computed steady-state flow field of the inner and outer fluids in the rest frame of the droplet due to mode~1. The flow field has vectorial symmetry and gives rise to the self-propulsion of the droplet in the laboratory frame.}
  \label{fig:Mode_1_Flow}
\end{figure}
\begin{figure}[!htb]
  \includegraphics[width = 1.0\linewidth]{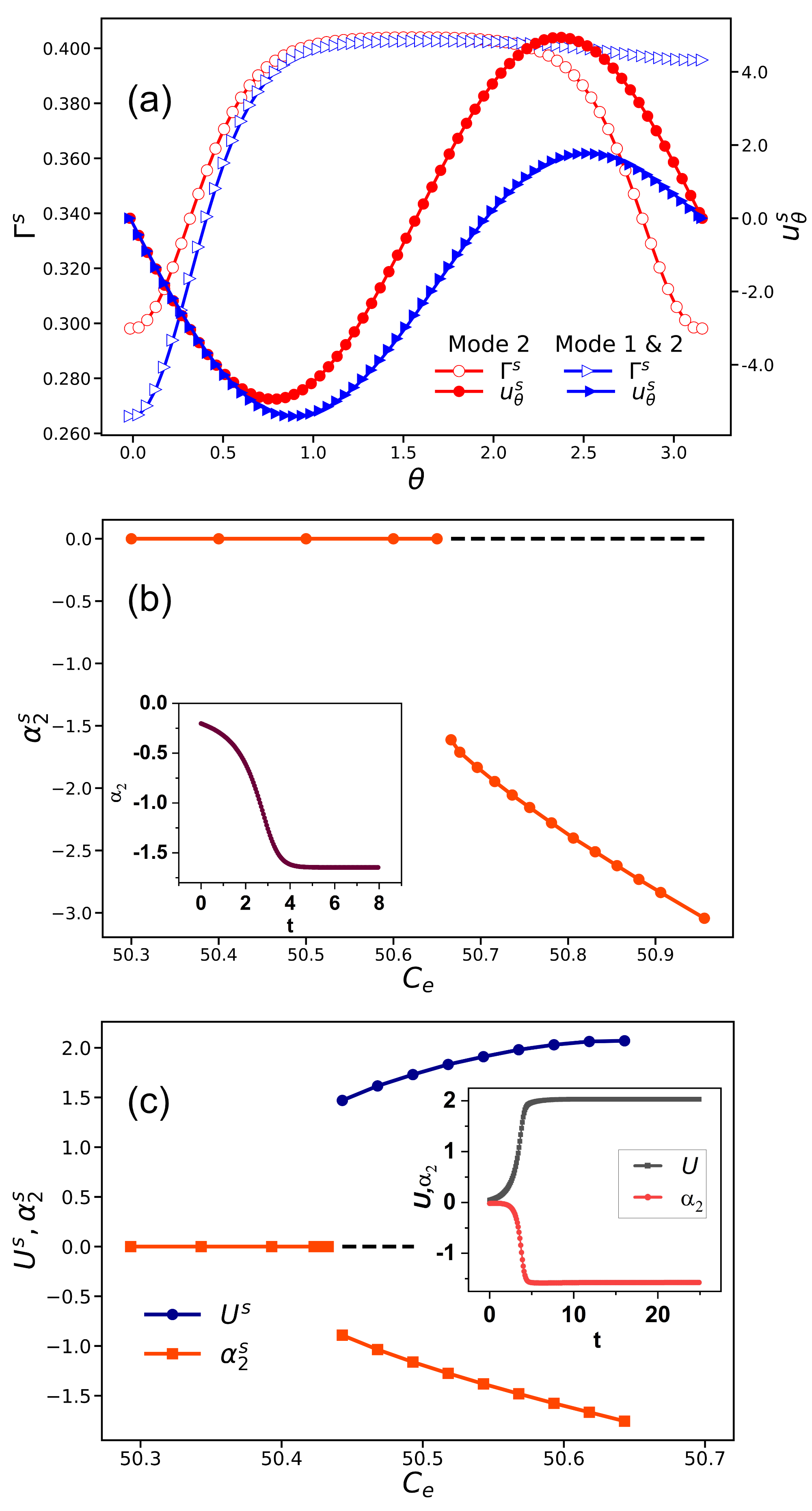}
  \caption{(a) The steady state surfactant concentration $\Gamma^s$ and tangential velocity profiles on the droplet interface for mode~2 at $C_e = 50.666$ (red graphs)  and combined modes at $C_e = 50.593$ (blue graphs). 
(b) The variation of the steady-state amplitude of the second mode with $C_e$ showing the onset of instability at 50.656. The inset in (b) depicts the time evolution of this amplitude above the threshold showing the saturation at long times. (c) The variation of the amplitudes with $C_e$ when both the first and second modes become unstable simultaneously. The inset in (c) depicts the time evolution of the amplitudes above the threshold.}
  \label{fig:Mode_12}
\end{figure}
\begin{figure}
  \includegraphics[width = 0.7\linewidth]{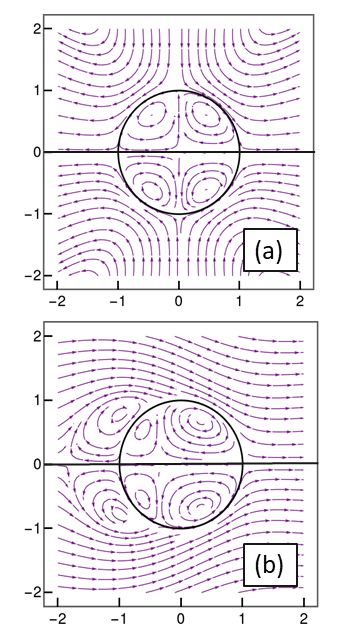}
  \caption{The numerically computed steady-state flow field of the inner and outer fluids in the rest frame of the droplet for (a) mode~2 and (b) the combination of mode~1 and mode~2.  The flow field for mode~2 has quadrupolar symmetry and gives rise to extensional flow with no net propulsion of the droplet. The combination of the two modes gives rise to a superposed flow pattern.}
  \label{fig:Mode_12_Flow}
\end{figure}
Fig.~\ref{fig:Mode_1}a shows the time evolution of the surfactant 
concentration profile and the
velocity field $u_\theta$ on the droplet surface for $C_e=50.493$ which is slightly above the threshold value.  Both $\Gamma$ and $u_\theta$ become increasingly non-uniform and tend to a steady state profile at long times.
The velocity profile $u_{\theta}$ peaks at $\theta = \pi/2$ whereas $\Gamma$
decreases at $\theta = 0$ and increases at $\theta = \pi$.
Accordingly, the droplet propulsion speed $U$ increases with time from zero to a steady state value (see inset of fig.~\ref{fig:Mode_1}b).
The steady-state droplet propulsion speed $U^s$ increases with increasing values of the driving parameter $C_e$ as shown in fig.~\ref{fig:Mode_1}b. Very close to the onset of the instability, $U^s$ grows as 
$(C_e- C_{e1})^{0.55}$ in our numerical model indicating that the instability corresponding to mode~1 has the signature of a supercritical bifurcation.
The steady-state velocity profiles of the inner and outer fluids in the droplet rest frame are shown in Fig.~\ref{fig:Mode_1_Flow} which are axially symmetric about the propulsion direction. The flow profile has a far-field velocity in the droplet rest frame, implying that the droplet has net propulsion in the laboratory frame.
     
The propulsion can be understood as follows. Small amplitude deviations in the surfactant concentration from its uniform value on the interface give rise to variations in the interfacial tension which generates a Marangoni flow
by spontaneous breaking of isotropic symmetry.  
The Marangoni flow requires that the surfactant concentration at the front is slightly greater than that at the rear end.  This induces a negative divergence of the in-plane flow field at the rear end and a positive divergence at the front end. According to our proposition, there is a greater probability of the emission of swollen micelles from the regions of negative divergence. These swollen micelles take away some surfactant molecules from the interface. This process tends to enhance the mode amplitude and is the source of activity in the system. On the other hand, the diffusion and advection processes tend to homogenize any non-uniformity in interfacial surfactant concentration. Above a critical value of the control parameter $C_e$, the droplet can maintain a lesser surfactant concentration at the trailing end and a higher concentration at the leading end and the resulting Marangoni flow. It is important to note that the total surfactant concentration in the outer fluid is the control parameter in our model as found in experimental studies.

\subsection{Mode 2 and combined Modes 1 \& 2}
Similarly, mode~2 corresponds to a surfactant distribution given by the Legendre polynomial of degree two. This mode has quadrupolar symmetry and gives rise to a steady extensile flow instead of the self-propulsion of the droplet. 
The threshold value of the driving parameter $C_e$ obtained from the linear stability analysis is given by eqn.~(\ref{eq:higher_threshold}) for $n=2$.
For the parameter values used in our model, the threshold value of $C_e$ for mode~2 is found to be 50.656, which is slightly greater than that for mode~1. It is found that mode~1 is always the first mode to get activated in our model.
The nonlinear analysis for mode 2 was performed for the driving parameter  
$C_e= 50.666$ which is slightly above the threshold value. The steady-state
profiles of surfactant distribution and tangential 
velocity are shown in fig~\ref{fig:Mode_12}a. 

For mode~2, since there is now negative surface divergence at both poles, the surfactant concentration at the poles is lower 
compared to the equatorial region and the distribution is symmetric about the equator. The tangential velocity profile at the droplet interface has the same magnitude but opposite direction about the equator.
Above the threshold value of the control
parameter, the flow amplitudes $\alpha_2$ grows from zero to
steady state values as shown in the inset of fig.~\ref{fig:Mode_12}b.
The variation of steady-state flow amplitude $\alpha_2^s$ with the control
parameter $C_e$ is shown in fig.~\ref{fig:Mode_12}b. The amplitude 
$\alpha_2^s$ increases from zero with a jump discontinuity at the onset of the instability indicating a subcritical bifurcation for this mode. The corresponding steady-state velocity profile in the droplet rest frame is shown in fig.~\ref{fig:Mode_12_Flow}a. 
The flow fields around the droplet have axial symmetry about the z-axis and mirror symmetry about the equatorial plane.

We also consider the excitation of these first two modes simultaneously in the system and studied the resulting steady-state surfactant distribution and flow profiles as shown in fig.~\ref{fig:Mode_12}a. In the combined mode, 
the magnitude of $u_{\theta }$ peaks closer to the rear stagnation point and the flow field is no longer mirror symmetric 
about $\theta = \pi/2$.
Above a threshold value of the control parameter, both the flow amplitudes grow from zero to steady state values as shown in the inset of fig.~\ref{fig:Mode_12}c. 
We find that the droplet swimming with the combined modes has the hydrodynamic signature of a pusher which has been established by recent experiments \cite{Hokmabad2021, Jin2021}. For the combined modes the steady-state velocity profile around the droplet for $C_e = 50.593$ is shown in fig.~\ref{fig:Mode_12_Flow}b. The combined mode gives rise to non-zero extensional flow contributions in addition to the propulsion even below the linear stability threshold for $\alpha_2$ as was observed for an isotropic phoretic particle \cite{Michelin2013}. We also observe that, when both modes are considered in the flow field, the first mode does not get activated for any value of $C_e$ if the initial perturbation to the surfactant distribution does not have the $P_1(\cos{\theta})$ term while the second mode gets activated even with zero initial perturbation corresponding to it.

\section{Conclusion}\label{Conclusion}
We propose a simple model for swimming active droplets suspended in a micellar solution. Our hydrodynamic model predicts the existence of a sharp instability threshold towards self-propulsion of the droplet in terms of total surfactant concentration in the micellar solution which agrees well with the experimental observations \cite{Herminghaus2014, Moerman2017}. 
Linear stability analysis was performed analytically to determine the instability threshold and full nonlinear equations were solved numerically to find the steady-state flow field in the fluids.
The theoretically calculated instability threshold agrees qualitatively with the experimentally determined values. Unlike the previous models which take into account only the gradient of surfactant concentration outside the droplet for the mechanism of self-propulsion, we show that self-propulsion could be achieved through interfacial processes alone with the spontaneous breaking of spherical symmetry. The direct emission of swollen micelles from the droplet interface is the dominant self-propulsion mechanism in our model. The experimental observation of a trail of filled micelles left behind by the moving droplet supports this mechanism. 

%

\end{document}